# Chlorophylls - natural solar cells

## Lorentz JÄNTSCHI[1,3*], Sorana D. BOLBOACĂ[2], Mugur C. BĂLAN[3], Radu E. SESTRAŞ[1]

[1] University of Agricultural Sciences and Veterinary Medicine Cluj-Napoca
[2] University of Medicine and Pharmacy "Iuliu Haţieganu" Cluj-Napoca
[3] Technical University of Cluj-Napoca
[*] Corresponding author: lori@academicdirect.org

**Abstract.** A molecular modeling study was conducted on a series of six natural occurring chlorophylls. Quantum chemistry calculated orbital energies were used to estimate frequency of transitions between occupied molecular orbital and unoccupied molecular orbital energy levels of chlorophyll molecules in vivo conditions in standard (ASTMG173) environmental conditions. Obtained results are in good agreement with energies necessary to fix the Magnesium atom by chlorophyll molecules and with occurrence of chlorophylls in living vegetal organisms.

## Introduction

Chlorophylls are small molecules containing a Magnesium atom responsible for conversion of the solar energy into chemical energy and represent the engine of any vegetal living organism. Vegetables spend the chemical energy produced by chlorophylls to construct complex organic compounds from inorganic compounds and/or simple organic compounds (Krause & Weis, 1991).

Only six different chlorophylls were identified in nature till nowadays (a - Conant & others, 1931a; b - Conant & others, 1931b; c1 & c2 - Strain & others, 1971; d - Miyashita & others, 1996; f - Chen & others, 2010). It is difficult to investigate the full role of chlorophylls in living cells because their action depend on many environmental parameters such as fixing proteins, surrounding water molecules, and solar spectrum (Maxwell & Johnson, 2000). Some authors suggests that when plants are exposed to light intensities in excess of those that can be utilized in photosynthetic electron transport, nonphotochemical dissipation of excitation energy is induced as a mechanism for photoprotection of photosystem II (Horton & others, 1996).

In the present paper two in-vitro approaches of molecular design were conducted on the series of six chlorophylls in order to relate the chemical properties with their natural occurrence.

## Material

Many studies in the literature were devoted to chlorophylls since 1863 (Fremy, 1863). The interest is sustained by the importance of the chlorophylls for propagation of life (Vines, 1879). Recent discover of chlorophyll f (Chen & others, 2010) may rise again questions about the diversity of chlorophylls among plants. In the present study were included into the analysis all naturally occurring proved chlorophylls.

The six known chlorophylls are depicted in Figure 1 (Magnesium atom: black; Nitrogen atoms: yellow; Oxygen atoms: light blue; Carbon atoms: red).

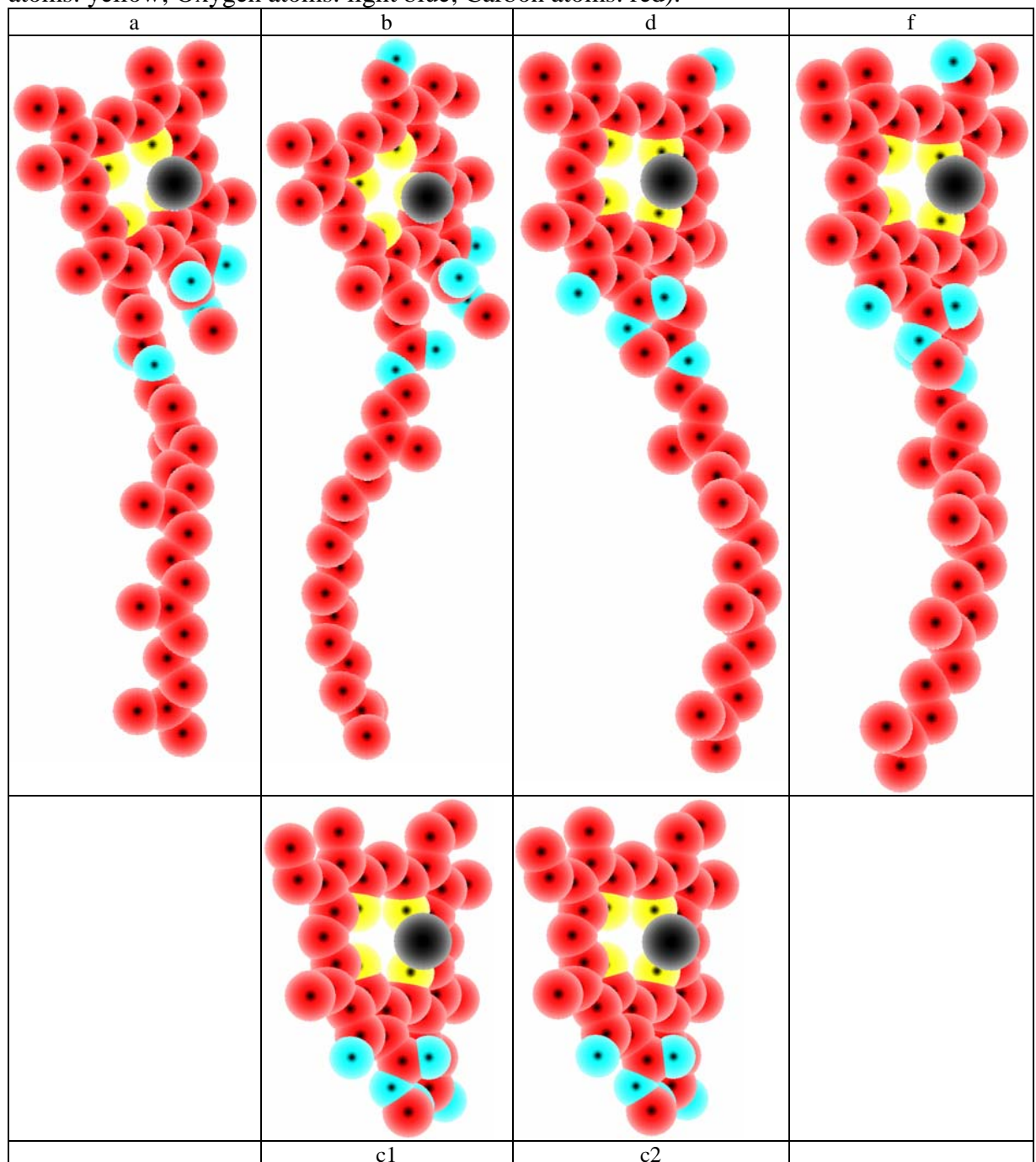

Fig. 1. The six natural occurring chlorophylls

By combining many reports on chlorophylls, we may conclude that their occurrence is not equal, and this diversity may be arisen from a long evolution and adaptation process. Table 1 gives a guess about the occurrence of chlorophylls.

Tab. 1.
Occurrence of chlorophylls

| Chlorophyll | a | b | d | f | c1 | c2 |
|---|---|---|---|---|---|---|
| ChemSpider ID | 16736115 | 16739843 | 16736116 | 2763140 | 391649 | 17229531 |
| Occurrence | universal | many plants | cyanobacteria | cyanobacteria | different algae | different algae |

**Methods**

*Method 1 (results in Table 2)*

- Geometries of the molecules were optimized for in-vitro conditions; OPLS total energies were recorded ($E_{wMg}$ in Table 2);
- Magnesium atoms were removed from molecules and bonds were reconstructed with hydrogen - following a suggestion given in two subsequent papers (Conant & Hyde, 1929; Conant & Hyde, 1930);
- Geometries of the molecules were optimized again for in-vitro conditions; OPLS total energies were recorded ($E_{w2H}$ in Table 2);
- Differences of the $E_{w2H}$ and $E_{wMg}$ were calculated ($\Delta E_{Mg}$ in Table 2);

*Method 2 (results in Table 3)*

- Chlorophylls were reconstructed from the systems obtained through method 1 by reconstructing Magnesium's bonds;
- Geometries of the chlorophylls were optimized for in-vitro conditions using OPLS molecular mechanics model till gradient of the optimization become less than 1‰; PM3 semi-empirical model were used to calculate molecular orbital next lowest states; from the pool of molecular orbital states were extracted the ones which correspond to OMO (occupied molecular orbital) - UMO (unoccupied molecular orbital) transitions in visible light known to be used by vegetal plants - domain expanded from 400-700 nm to 310-700 nm as suggested in (Chen & Blankenship, DOI);
- ASTMG173 (American Society for Testing and Materials G173: Reference Solar Spectral Irradiance, developed by subcommittee G03.09, book of standards volume 14.04) modeled using the SMARTS v.2.9.2 (Gueymard, 2001; Gueymard, 2004) were used to obtain solar intensities corresponding to each transition - results in Table 3;
- For every excited UMO state were assigned an intensity (by adding solar intensities if more than one transition ends to the UMO state);
- For HOMO (highest occupied molecular orbital) state were calculated the electron density as sum between 1 (it's fundamental state) and ratio of solar intensity promoting the electron to this state from total absorbed solar intensity (from 310-700 nm range);
- For UMO excited states proportions were calculated from assigned solar intensities of possible transitions; Shannon entropies (Shannon, 1948) of the excited UMO states were calculated from these proportions;
- Finally, a solar recovery index were calculated multiplying HOMO state electron density with Shannon entropies of the excited UMO states;

**Results**

The energies of Magnesium's removal are given in Table 2. It should be noticed that as energy ($\Delta E_{Mg}$) increases the bonding is stronger and is expected that electron excitations to be lower.

Tab. 2.
Magnesium removal method (Method 1)

| Chlorophyll | a | b | d | f | c1 | c2 |
|---|---|---|---|---|---|---|
| Formula | $C_{55}H_{72}MgN_4O_5$ | $C_{55}H_{70}MgN_4O_6$ | $C_{54}H_{70}MgN_4O_6$ | $C_{55}H_{74}MgN_4O_6$ | $C_{35}H_{30}MgN_4O_5$ | $C_{35}H_{28}MgN_4O_5$ |
| $E_{w2H}$ [kcal/mol] | 79.387 | 65.679 | 78.419 | 65.635 | 90.072 | 87.323 |
| $E_{wMg}$ [kcal/mol] | 119.276 | 103.629 | 220.501 | 204.804 | 185.548 | 182.707 |
| $\Delta E_{Mg}$ [kcal/mol] | 39.89 | 37.95 | 142.08 | 139.17 | 95.48 | 95.38 |
| $\log(\Delta E_{Mg})$ | 1.60 | 1.58 | 2.15 | 2.14 | 1.98 | 1.98 |

Table 3 below contains the analysis of molecular orbital states, conducted for the electron transitions in 310..700 nm range. It should be noted that for all chlorophylls the highest two occupied states contains one electron each, in absence of the light excitation. Thus, one electron can be promoted from before highest occupied state to highest occupied state if the light meets the energy requirements (in Table 3 are colored in blue these special cases). Since are more than one possible transition, the absorbed light will be used to promote the electron from first state with energy below. Thus, transitions will split the white light in bands (Band column in Table 3). For solar light, every band has a known standard energy, which was computed from ASTMG173 reference data (SI-B column). For the list of unoccupied molecular orbital states (UMO column) were cumulated solar intensities exciting electrons to these states (SI-U column) as well as their partition (SI-U% column).

Tab. 3.
Next lowest state transitions (Method 2)

| Chlorophyll | State(eV)/Type(U/O) | | Energy(eV) | Band | SI-B | UMO | SI-U | SI-U% | $H_1$ |
|---|---|---|---|---|---|---|---|---|---|
| a | -3.70/O | -1.34/U | 2.36 | [379.8,409.0) | 17.7 | -0.43 | 49.9 | 24.06 | 0.343 |
| $\rho_e$=1.432 | -3.70/O | -1.09/U | 2.61 | [409.0, 475.1) | 73.4 | -0.67 | 66.5 | 32.03 | 0.365 |
| $\Sigma H_1$=1.285 | -3.70/O | -0.67/U | 3.03 | [475.1, 525.9) | 66.5 | -1.09 | 73.4 | 35.38 | 0.368 |
| $\rho_e\Sigma H_1$=1.84 | -3.70/O | -0.43/U | 3.27 | [525.9, 562.9) | 49.9 | -1.34 | 17.7 | 8.530 | 0.210 |
| | -5.90/O | -3.70/O | 2.20 | [562.9, 700.0) | 175.2 | | | | |
| b | -3.87/O | -0.28/U | 3.59 | [345.5, 406.6) | 29.0 | -0.28 | 29.0 | 8.01 | 0.202 |
| $\rho_e$=1.082 | -3.87/O | -0.82/U | 3.05 | [406.6, 434.9) | 25.8 | -0.82 | 25.8 | 7.11 | 0.188 |
| $\Sigma H_1$=1.426 | -3.87/O | -1.02/U | 2.85 | [434.9, 538.9) | 134.7 | -1.02 | 134.7 | 37.16 | 0.368 |
| $\rho_e\Sigma H_1$=1.54 | -6.17/O | -3.87/O | 2.30 | [538.9, 563.6) | 33.4 | -1.67 | 64.2 | 17.70 | 0.307 |
| | -3.87/O | -1.67/U | 2.20 | [563.6, 612.3) | 64.2 | -1.84 | 108.8 | 30.02 | 0.361 |
| | -3.87/O | -1.84/U | 2.03 | [612.3, 700.0) | 108.8 | | | | |
| c1 | -4.00/O | -0.34/U | 3.66 | [339.2,420.5) | 43.5 | -0.34 | 43.5 | 12.47 | 0.260 |
| $\rho_e$=1.117 | -4.00/O | -1.05/U | 2.95 | [420.5,427.7) | 6.5 | -1.05 | 6.5 | 1.87 | 0.075 |
| $\Sigma H_1$=1.190 | -4.00/O | -1.10/U | 2.90 | [427.7,486.1) | 70.4 | -1.10 | 70.4 | 20.20 | 0.323 |
| $\rho_e\Sigma H_1$=1.33 | -4.00/O | -1.45/U | 2.55 | [486.1,509.7) | 30.2 | -1.45 | 30.2 | 8.65 | 0.212 |
| | -4.00/O | -1.56/U | 2.43 | [509.7,660.3) | 198.1 | -1.56 | 198.1 | 56.81 | 0.321 |
| | -5.88/O | -4.00/O | 1.88 | [660.3,700.0) | 47.4 | | | | |
| c2 | -4.00/O | -0.36/U | 3.64 | [341.0,424.5) | 47.1 | -0.36 | 47.1 | 13.54 | 0.271 |
| $\rho_e$=1.120 | -4.00/O | -1.08/U | 2.92 | [424.5,439.2) | 13.5 | -1.08 | 13.5 | 3.88 | 0.126 |
| $\Sigma H_1$=1.229 | -4.00/O | -1.18/U | 2.82 | [439.2,490.5) | 64.7 | -1.18 | 64.7 | 18.60 | 0.313 |
| $\rho_e\Sigma H_1$=1.38 | -4.00/O | -1.47/U | 2.53 | [490.5,510.7) | 26.6 | -1.47 | 26.6 | 7.64 | 0.196 |
| | -4.00/O | -1.57/U | 2.43 | [510.7,659.6) | 195.9 | -1.57 | 195.9 | 56.34 | 0.323 |
| | -5.88/O | -4.00/O | 1.88 | [659.6,700.0) | 48.7 | | | | |
| d | -5.40/O | -1.41/U | 3.99 | [311.1,320.4) | 0.6 | -0.14 | 5.8 | 1.44 | 0.061 |
| $\rho_e$=1.000 | -4.01/O | -0.14/U | 3.87 | [320.4,346.5) | 5.8 | -0.46 | 9.7 | 2.43 | 0.090 |
| $\Sigma H_1$=1.382 | -5.40/O | -1.82/U | 3.58 | [346.5,350.1) | 0.9 | -0.70 | 39.6 | 9.89 | 0.229 |
| $\rho_e\Sigma H_1$=1.38 | -4.01/O | -0.46/U | 3.54 | [350.1,375.6) | 9.7 | -1.12 | 57.0 | 14.23 | 0.278 |
| | -4.01/O | -0.70/U | 3.30 | [375.6,430.1) | 39.6 | -1.41 | 119.1 | 29.73 | 0.361 |
| | -4.01/O | -1.12/U | 2.88 | [430.1,478.2) | 57.0 | -1.82 | 169.4 | 42.29 | 0.364 |
| | -4.01/O | -1.41/U | 2.59 | [478.2,567.2) | 118.4 | | | | |
| | -4.01/O | -1.82/U | 2.19 | [567.2,700.0) | 168.5 | | | | |
| f | -5.22/O | -1.28/U | 3.95 | [314.3,333.1) | 2.7 | -0.11 | 3.6 | 0.89 | 0.042 |
| $\rho_e$=1.000 | -3.84/O | -0.11/U | 3.72 | [333.1,347.4) | 3.6 | -0.47 | 12.4 | 3.08 | 0.107 |
| $\Sigma H_1$=1.210 | -5.22/O | -1.65/U | 3.57 | [347.4,368.8) | 7.5 | -0.70 | 95.6 | 23.80 | 0.342 |
| $\rho_e\Sigma H_1$=1.21 | -3.84/O | -0.47/U | 3.36 | [368.8,395.1) | 12.4 | -1.28 | 114.3 | 28.44 | 0.358 |
| | -3.84/O | -0.70/U | 3.14 | [395.1,484.0) | 95.6 | -1.65 | 176.0 | 43.79 | 0.362 |
| | -3.84/O | -1.28/U | 2.56 | [484.0,567.2) | 111.6 | | | | |
| | -3.84/O | -1.65/U | 2.19 | [567.2,700.0) | 168.5 | | | | |

Since transitions are delay based, entropies of molecular orbital states usage by the solar light it's an indicator of unoccupied molecular orbital states usages efficiency ($H_1$ column calculated from SI-U% column). Other important indicator is given by the relative population of the HOMO state with electrons. For all cases fundamental HOMO state contains one electron. In some cases (lines in blue in Table 3), a part from solar radiation are spent to promote the second electron at HOMO. For these cases, the ratio of the solar light intensity spent for promotion of this electron from the entire solar light intensity in the 310..700 nm range (405.2 $Wm^{-2}$) give the population of the HOMO with the second electron ($\rho_e$ values in Table 3 cumulates the availability of electrons on HOMO state). By multiplying unoccupied molecular orbital states usages efficiency ($\Sigma H_1$ estimator in Table 3) with highest occupied molecular orbital electron availabilities ($\rho_e$ estimator in Table 3) were obtained a clue (relative scale) of solar energy conversion efficiency of chlorophylls.

**Discussions**

As were noted before, it is very difficult to measure or even estimate the in vivo action of chlorophylls under solar light, because their action depend on many environmental parameters and our measurements or models can be easily affected by not taking them into account. Two simple approaches were conducted in order to relate the natural occurrence with quantum chemical calculations of in-vitro structures of chlorophylls, by using the reasoning that if these differences exist in occurrence, it should be observed at molecular level as well.

The results obtained in Table 2 suggests that chlorophyll b should be observed more often than chlorophyll a (Mg binding energy lower in b than in a) and this may be an acceptable result. The Mg binding energies of c1 and c2 chlorophylls are significantly lower than Mg binding energies of d and f chlorophylls and suggest that c1 and c2 chlorophylls are more often occurring than d and f chlorophylls, and this is not an acceptable result. The only possible conclusion which can be drawn is that the Mg binding energies has no role on propagation of the chlorophylls between living plants.

The second method, more accurate designed (Table 3) takes into account the transitions between molecular energy levels. The results of second method are in better agreement with observed natural occurrences (Table 4).

Tab.4
Natural and estimated occurrences of chlorophylls

| Chlorophyll | Occurrence | $\rho_{e,HOMO} \cdot \Sigma H_{1,UMO}$ | Estimated occurrence | Observations |
|---|---|---|---|---|
| a | universal | 1.84 | **** | Probability to be different from the mean of {1.54, 1.38, 1.21, 1.33, 1.38} is over 99.9% |
| b | many plants | 1.54 | *** | Probability to be different from the mean of {1.38, 1.21, 1.33, 1.38} is over 99.4% |
| d | cyanobacteria | 1.38 | ** | Probability to be different one to each other is less than 22% |
| c1 | different algae | 1.33 | ** | |
| c2 | different algae | 1.38 | ** | |
| f | cyanobacteria | 1.21 | * | Probability to be different from the mean of {1.84, 1.54, 1.38, 1.33, 1.38} is over 96.1% |

As Table 4 showed, solar energy conversion efficiency of chlorophylls expressed by $\rho_{e,HOMO} \cdot \Sigma H_{1,UMO}$ terms agrees with its observed natural occurrences. The advantage of solar energy conversion efficiency of chlorophylls is given by the fact that these are expressed on a numerical scale which may allow comparisons. Thus, if Student t-test (Student, 1908) are involved to compare the individual marginal values with the mean of the all other values, three major groups of different occurrences are created (estimated occurrence column in Table 4).

## Conclusions

Two different quantum chemistry methods were involved in order to relate natural occurrence of chlorophylls with molecular level chemical parameters. The results obtained from first method of analysis showed that the binding energy of Magnesium have no direct effect on efficiency of solar energy conversion by chlorophylls. Opposite, results obtained conducting second method of analysis showed that the in-vitro next lowest molecular orbital states of chlorophylls are in good agreement with the observed natural occurrences of chlorophylls.

## Acknowledgments

The study was supported by POSDRU/89/1.5/S/62371 through a postdoctoral fellowship for L. Jäntschi.